%% file: main.tex
\documentclass[runningheads]{llncs}
\usepackage{graphicx}
\usepackage{graphics}
\usepackage{csquotes}

\usepackage[hyphens]{url}
\usepackage{hyperref}
\usepackage[hyphenbreaks]{breakurl}

\usepackage{listings}
\usepackage{framed}
\usepackage{textpos}
\usepackage[space]{cite}

\lstdefinelanguage{JavaScript}{
	keywords={},
	keywordstyle=\color{bluegrey}\bfseries,
	morekeywords=[2]{attributes, class, classend, do, empty, endif, endwhile, fail, function, functionend, if, implements, in, inherit, inout, not, of, operations, out, return, set, then, types, while, use},
	keywordstyle={[2]\color{violet}\bfseries},
	otherkeywords={@param, @returns, @author, @type, @link, @see},
	sensitive=false,
	morecomment=[l][\color{lstgreen}]{//},
	morecomment=[s][\color{lstgreen}]{/*}{*/},
	morecomment=[s][\color{javadoc}]{/**}{*/},
	morestring=[b]',
	morestring=[b]"
}

\begin{document}
\title{Configurable Per-Query Data Minimization for Privacy-Compliant Web APIs}
\titlerunning{Per-Query Data Minimization for Privacy-Compliant Web APIs}
\author{Frank Pallas\orcidID{0000-0002-5543-0265} \and
David Hartmann\orcidID{0000-0001-9745-5287} \and \\
Paul Heinrich\orcidID{0000-0001-9068-3260} \and
Josefine Kipke\orcidID{0000-0002-6782-8393} \and \\ 
Elias Grünewald\orcidID{0000-0001-9076-9240}
}
\authorrunning{F. Pallas et al.}

\institute{TU Berlin, Information Systems Engineering, Berlin, Germany\\
\email{\{fp,dh,ph,jk,eg\}@ise.tu-berlin.de}}

\maketitle              %
\begin{abstract}
The purpose of regulatory data minimization obligations is to limit personal data to the absolute minimum necessary for a given context. Beyond the initial data collection, storage, and processing, data minimization is also required for subsequent data releases, as it is the case when data are provided using query-capable Web APIs. Data-providing Web APIs, however, typically lack sophisticated data minimization features, leaving the task open to manual and all too often missing implementations.
In this paper, we address the problem of data minimization for data-providing, query-capable Web APIs. Based on a careful analysis of functional and non-functional requirements, we introduce \textsc{Janus}, an easy-to-use, highly configurable solution for implementing legally compliant data minimization in GraphQL Web APIs. \textsc{Janus} provides a rich set of information reduction functionalities that can be configured for different client roles accessing the API. We present a technical proof-of-concept along with experimental measurements that indicate reasonable overheads. \textsc{Janus} is thus a practical solution for implementing GraphQL APIs in line with the regulatory principle of data minimization.

\end{abstract}

\begin{textblock*}{1.2\textwidth}(-1.5cm,-17.5cm) %
    \begin{center}
    \begin{framed}
        \textit{preprint version (2022-03-18) This version of the contribution has been accepted for publication after peer review but is not the Version of Record and does not reflect post-acceptance improvements, or any corrections.\\
		\textbf{22nd International Conference on Web Engineering (ICWE 2022)}\\
		To be published in the SpringerLink Digital Library: https://link.springer.com/conference/icwe}
    \end{framed}
        
    \end{center}
    \end{textblock*}

\input{00_contents}

 \bibliographystyle{splncs04}
  \bibliography{references}

\end{document}

%% file: 00_contents.tex
\section{Introduction}

Data minimization is one of the core principles of privacy regulations such as the GDPR. Basically, it requires to limit personal data to the absolute minimum necessary in a given context. Beyond collection, storage and processing of personal data, this minimization obligation also applies to subsequent data releases. Any such release of personal data -- between different departments of an organization or to external parties -- must thus also be confined to the absolute minimum required by the particular recipient. Depending on the usecase and the client role, this can require to pseudonymize data, to strip off certain sensitive attributes, or to apply information reduction methods such as generalization or noising to avoid re-identification. %

Real-world usecases involving such releases of personal data today typically employ query-capable Web APIs following paradigms such as REST \cite{fielding2000representational} or GraphQL \cite{brito2019}. Existing technology stacks broadly used in industry to implement such APIs do, however, so far not provide the means necessary for implementing above-mentioned data minimization techniques in a developer-friendly, coherent, and reliable fashion easily adoptable to the different minimization requirements applicable for different usecases and data-requesting parties (or roles). Beyond fundamental mechanisms for access control, data controllers providing personal data via Web APIs are thus currently left without proper technical support for %
meeting regulatory requirements. The only alternative currently lies in individually implemented %
external wrapper components, which raise significant efforts %
and are error-prone.

We herein close this gap by introducing the concept of per-query role-based data minimization for  data-providing Web APIs. We identify a set of functional and non-functional requirements that must be met by a technical mechanism in order to fulfill regulatory requirements and be practically applicable. On this basis, we present %
\textsc{Janus}, a ready-to-use extension to one of the most widely used software stacks for building GraphQL APIs -- Apollo -- that facilitates low-effort integration of a broad variety of data reduction techniques. %
All components %
are provided %
under an open source license in publicly available repositories.

The remainder of this paper is structured as follows: Section~\ref{sec:background} provides relevant preliminaries on %
legal requirements for data minimization and on respective technical measures for implementing it in practice. A motivating and illustrative scenario is also provided here. %
On this basis, we distill functional and non-functional requirements in section~\ref{sec:requirements} and elaborate on the integration approach, architecture, provided functionality, and performance assessment of our prototypical implementation in section~\ref{sec:implementation}. %
Limitations of our approach, pathways for future work and a conclusion are provided in section~\ref{sec:limitations}. %

\section{Preliminaries} \label{sec:background}

In the following, we set out the necessary preliminaries to contextualize our approach in the light of legal and technical givens and %
provide an illustrative scenario mitivating and guiding our subsequent considerations.%

\subsection{Regulatory Background}\label{sec:regulations}

As briefly touched above, %
data minimization is a core concept of modern privacy regulations. %
The GDPR \cite{gdpr} can be taken as a representative for comparably structured legislations such as California's CCPA or China's PIPL here: In Art. 5 (1c), it requires that \enquote{personal data shall be [\ldots] limited to what is necessary in relation to the purposes for which they are processed}. %
Noteworthily, data minimization must not only be applied to the collection of personal data but also for their processing and for providing access to them \cite{edpb-dpbd-2019,gdpr}.

In practice, this can be done in two different veins: First, the amount of data can be minimized. As the European Data Protection Board points out \cite{edpb-dpbd-2019}, this refers to the \enquote{quantitative and qualitative} amount, thus including \enquote{the volume of personal data, as well as the types, categories and level of detail}. Insofar, the data minimization principle requires to remove as many attributes of the data as possible and to limit the level of detail for the remaining ones %
to the absolute minimum required in a given context. %

Second, the minimization principle does not %
require to minimize the amount of %
data in general but only that of \emph{personal} data. Another viable approach is thus to anonymize (``de-personalize'') initially personal data. In so doing, the mere removal of explicit identifiers such as names is, however, typically not sufficient due to re-identification risks. %
Different information reduction techniques (see section \ref{sec:anon-reduc}) can reduce these risks and %
ultimately render data non-personal.%

For both approaches, the required level of information reduction and %
de-personalization cannot be determined universally but must %
be assessed on a per-case basis, taking into account factors such as the nature and scope of the data, the context it is to be processed in, etc. \cite{finck-pallas-identified}. This particularly also includes the distinction between different data recipients: releasing data to an academic research group will, for instance, typically require less strict minimization than providing it to an international insurance company or even the general public. %

\subsection{Information Reduction \& Anonymization}\label{sec:anon-reduc}

From the technical perspective, information reduction and anonymi\-za\-tion can take place in different forms \cite{marques20-anon,anon-tech-survey,gruschka2018privacy}: \emph{Attribute suppression} means to completely remove certain attributes (such as an explicit identifier or a particular characteristic) from a data point. \emph{Generalization}, in turn, reduces the level of detail at which an attribute is included. Typical examples here comprise the replacement of detailed dates-of-birth with more general year-of-birth ranges, blinding %
digits from a ZIP code (sometimes considered as a separate technique of \emph{character replacement}), and so forth. \emph{Hashing} herein refers to substituting a value with the result of a (basically) non-revertible hash-function, retaining uniqueness and validatability without revealing the underlying plain-text data.\footnote{Hashing is thus often considered as a particular form of pseudonymization when applied to identifiers. A substantial reversion risk may, however, still exist %
-- for a detailed discussion, see \cite{finck-pallas-identified}.}

Beyond these mechanisms, ano\-ny\-mi\-ty measures such as $k$-anonymity \cite{sweeney_k-anonymity_2002}, $\ell$-diversity \cite{machanavajjhala_l-diversity_2007}, or $t$-closeness \cite{li_t-closeness_2007} were introduced to guarantee certain levels of non-identifiability within a dataset. However, these measures as well as the techniques and algorithms for implementing them are targeted at (rather) static datasets that are to be released only once or on rather infrequent occasions. In the context of query-capable APIs delivering continuously changing data, such anonymization schemes cannot be reasonably applied.

For such contexts, different forms of \emph{noising} (sometimes also referred to as \emph{perturbation}) are thus proposed. Advanced approaches of \enquote{differential privacy} \cite{dwork2008differential} here provide statistical guarantees but can only be applied to aggregating queries such as sum, count, etc. and are thus limited to a particular class of usecases. When data are needed in non-aggregated form, in turn, noising is typically done in a way that makes individual values \enquote{less accurate whilst retaining the overall  distribution} \cite{art29-anon-2014}, e.g. through in-/decreasing numerical values according to typical probability distributions with the level of noise depending on \enquote{the level of information [detail] required and the impact on individuals’ privacy} \cite{art29-anon-2014}.

\subsection{Data-Providing Web APIs and Data Minimization}
The broad reception of these and further %
techniques and their importance for achieving regulatory compliance notwithstanding, established and easily re-usable technical implementations are currently missing. Where available at all, respective programming libraries so far only provide the mathematical or algorithmical core functionality while lacking coherent and low-effort integration into current application architectures and, in particular, programming stacks for building data-providing Web APIs.%

Such APIs today mostly follow one of the two paradigmatic approaches of REST and GraphQL. Of these, GraphQL provides significant benefits over REST in matters such as request-response-latencies or the amount of data to be transferred in real-world usecases \cite{vogel-graphql-perf,wittern_empirical_2019}. Together with its capabilities for client-specified queries \cite{brito2019}, this increasingly makes GraphQL the paradigm of choice for implementing data-providing Web APIs, especially when relevant data structures become more complex and when different parties only need certain subsets of the data. We therefore focus on GraphQL herein.

In service-oriented architectures, such APIs are used by external parties (such as the users of a given app or third parties) as well as by internal ones. As soon as this involves personal data, the minimization principle comes into play, requiring the provided data to be limited to the amount absolutely necessary for the respective party and/or role. However, technical tools for doing so in established GraphQL programming stacks are rare. Existing approaches such as \textit{GraphQL Shield} \cite{zavadal2021} or \textit{GraphQL RBAC} \cite{cannergraphql-rbac_2021} only implement simple permission layers and do not support the implementation of further information reduction and anonymization functionalities. %

Altogether, we thus know a broad variety of fundamental information reduction and anonymization techniques aimed at the data minimization principle. When used in proper combination, they may, depending on the type of data, the context, and the party receiving the data, even allow to render data non-personal from the regulatory perspective. At the same time, the practical application of these techniques in real-world Web APIs is -- like for other privacy / data protection principles and technologies \cite{gruenewald2021tira,pallas2020pbac} -- hindered by a lack of easily adoptable technical solutions that smoothly integrate into established technology stacks and development practices \cite{kostova2020privacy}. %

\subsection{Illustrative Scenario} \label{sec:application}

To guide and illustrate our subsequent considerations, we assume the exemplary scenario of a period tracking app implementing %
a common architecture %
with a smartphone- or web-application sending and retrieving data to/from a Web API, which, in turn, stores and retrieves data in/from a backend database.

Basically, such period tracking apps provide valuable insights for their users regarding %
estimated pain, contraception and ovulation. At the same time, however, the web API may also be used for sharing menstrual data %
with other parties %
to generate additional benefits: With sufficiently minimized and/or de-personalized menstrual data being queryable from the API, a scientific research group would, for instance, be able to gain new insights on the relationship between health and periods. Public health programs, in turn, would be able to better recognize and counteract existing challenges related to menstrual hygiene (including, e.g., a lack of infrastructure available) based on such data 
\cite{ijerph17082633}. Internal processes of app development %
might also benefit from appropriately minimized usage data and, last but not least, even users themselves could profit from queries like \enquote{how severe is my pain compared to other users of the same age cohort} being facilitated by the API and therefore available in the app.

In all these and many further usecases, the sharing of –- sufficiently minimized -- period data with parties beyond the data subjects themselves proposes noteworthy societal or individual benefits. On the other hand, given the sensitive nature of such data, the technical mechanisms for doing so must also be reliable and ideally not implemented individually in an ad-hoc fashion, motivating the development of a re-usable component that can be easily integrated into existing Web API frameworks. %
Any such component must fulfill several functional and non-functional requirements which shall be laid out below.

\section{Requirements} \label{sec:requirements}

In line with other endeavors of practical privacy engineering (such as \cite{pallas2020pbac,gruenewald2021,gruenewald2021tira}), 
we formulate a set of functional and non-functional requirements that need to be fulfilled. %
Functional requirements here refer to the core functionality that needs to be provided while non-functional requirements address the practical applicability in real-world technology stacks and architectures.

\subsection{Functional Requirements}

\paragraph{Attribute-Level Role-Based Access Control (FR1):} The first step towards data minimization depending on different parties or roles %
accessing a GraphQL API is to restrict access to single attributes of data items depending on the accessing party. In our illustrative scenario, an external academic research team might, for instance, be allowed to access detailed menstrual data but only without identifiers such as names etc., while internal account management might access these identifiers but not the sensitive attributes like menstrual cycles or pain. Any solution must therefore implement access control on a per-attribute level. To integrate well with broadly established practices in access management and control in general, %
doing so on the basis of roles 
appears most appropriate.

\paragraph{Attribute-Level Role-Based Information Reduction (FR2):} Besides the mere blocking of access to single attributes, it must be possible %
to implement data minimization through applying different forms of information reduction to different attributes, again according to different roles performing data access. In the just-mentioned example, the detailed data provided to an academic research team might have to be subject to generalization of age cohorts to meet regulatory requirements while an internal product improvement team may only see noised usage patterns. %

\paragraph{Rich and Extendible Set of Information Reduction Methods (FR3):} As effective data minimization %
is subject to highly case-specific requirements that must be met for achieving regulatory compliance, a diversity of different information reduction techniques can be necessary, ranging from numerical categorization over character replacements to different approaches of statistical noising. This calls for a rich set of such functionalities -- covering numerical and non-numerical values -- to be built upon when defining case- and role-specific information reduction schemes. Ideally, this set should be easily extensible in onward development to accommodate additionally identified information reduction needs.%

\paragraph{Configurability (FR4):} In addition, any technical solution must be highly configurable and allow for adjustable %
levels of information reduction to satisfy different regulatory requirements while meeting certain accuracy constraints \cite{ghinita2009framework}. %

\subsection{Nonfunctional Requirements}

\paragraph{Low Integration Overhead (NFR1):} Smooth and low-effort integration %
into at least one software stack widely used in practice for implementing data-providing Web APIs fosters practical applicability and viability. At the same time, the connection to an externally maintained role-definition and authentication subsystem -- e.g., via widely-used JSON Web Tokens (JWTs) -- is a necessary precondition for being interoperable with already existing system architectures. %

\paragraph{Reusability (NFR2):} %
Reusability in a broad variety of application architectures, thorough documentation, and public availability foster software artifacts' practical adoption. This also comprises the availability under an open source licence for commercial use via common code repositories 
and distribution via package managers. The latter, eventually, introduce quality assurance, e.g., through code linting, automatic update mechanisms, %
and security alerts.%

\paragraph{Reasonable Performance Overhead (NFR3):} Article of the 25 GDPR states that technical measures for materializing privacy principles must be applied \enquote{depending on the cost of implementation}. %
Any solution must thus not introduce disproportional performance overheads. %
The overheads must, in turn, be experimentally determined in realistic settings and with different configurations to demonstrate the practical viability. %

\section{Approach \& Implementation} \label{sec:implementation}
To fulfill these requirements, we introduce the concept of per-query role-based data minimization in Web APIs and provide a ready-to-use implementation for Apollo, one of the most-widely used software stacks for implementing GraphQL APIs. Given its broad adoption in practice, building upon Apollo is a promising starting point for keeping integration overhead low for practitioners (see \emph{NFR1}). Besides mere adoption, Apollo also provides a mature framework for implementing custom extensions (for details, see below) and thus allows to integrate our intended functionality in a modular and low-effort fashion, supporting the fulfillment of \emph{NFR1} even further. 

Following this fundamental choice of Apollo as our target platform, the integration approach, implementation details, practical usage, and experimentally determined overheads of our prototypical implementation \textsc{Janus} shall be laid out below.

\subsection{Architecture Integration} \label{sec:architectureIntegration}
As delineated in \emph{FR1 and FR2}, \textsc{Janus} must provide functionalities for attribute-level access control and information reduction. Basically, these 
can be implemented in GraphQL using either a middleware- or a schema-directive-driven approach. Both can extend Apollo's integrated resolver functionality to first query a data field and then subject it to further processing, thus smoothly integrating into Apollo's general design and processing flow (see \emph{NFR1}). However, they function in a significantly different manner: In the middleware-driven approach, the resolver can be provided with middleware functions that execute any logic before and after the resolution of the field. %
Furthermore, middleware functions can call subsequent ones when passing the execution and modify the request and response object at any time, thus possibly creating an onion-like resolving flow. 

GraphQL schema directives, in turn, define post-processing steps to be executed on already resolved data fields before returning them through the API. Configured directly in the schema by adding the directive behind the targeted data field, directives can be used on specified fields or types of schemas. It is also possible to define a directive pipeline with subsequent directives. Compared with the middleware approach, integrating additional functionality via schema directives is simpler and provides more clarity (expectably leading to less integration overhead for developers, see \emph{NFR1}) while still allowing to implement a sufficient level of logical complexity through combining multiple schema directives. We thus chose the approach of schema directives for integrating our functionality into the Apollo stack at runtime, employing the \texttt{SchemaDirectiveVisitor} class provided by the \textit{graphql-tools}\footnote{\url{https://github.com/ardatan/graphql-tools}} library.

\begin{figure}[t]
    \centering
    \includegraphics[width=1.0\linewidth]{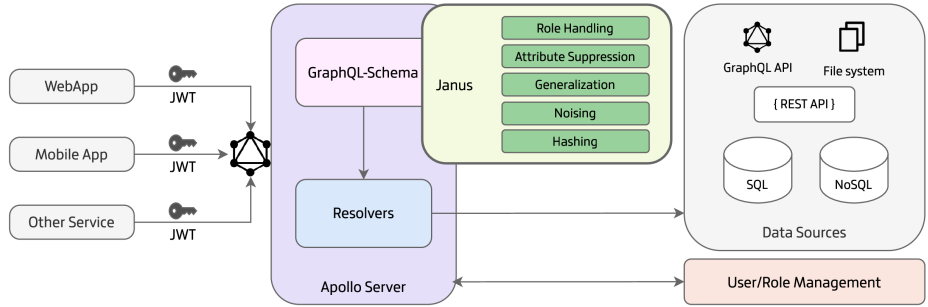}
    \caption{Apollo server architecture and \textsc{Janus} integration.}
    \label{fig:puzzel}
\end{figure}

Both, access control and information reduction are to be implemented on the basis of roles (see \emph{FR1 and FR2}). We thus need a mechanism for managing users, assigning them to roles, letting them log in and then make requests on the basis of these roles. On the server side, in turn, above-mentioned schema directives must be defined depending on these roles and requests must be processed accordingly. To ensure integratability with a wide variety of pre-existing systems and architectures here (\emph{NFR1}), we opted to keep role management, authentication, etc. external to our solution and to base our functionality on externally maintained roles provided via JSON Web Tokens. \textsc{Janus} simply needs the role parameter to be passed for mapping the role to a set of schema directives to be applied. Figure \ref{fig:puzzel} depicts the resulting general architecture.

\subsection{Implementation}
Given this general architecture and integration approach, \textsc{Janus'} functionality is implemented in two separate modules with well-distinguished scopes: First, a module is needed that provides a rich set of information reduction methods for various data types (\emph{FR3}). These are provided in the package \textit{janus-value-anonymizer}, which is publicly available under an open source license on Github as well as in the npm package managing system (thus meeting \emph{NFR2}).\footnote{\url{https://github.com/PrivacyEngineering/janus-value-anonymizer}}%
Possible extensions with additional information reduction methods (as required in \emph{FR3}) can easily be introduced in this module. In addition, this module can also be used in other contexts than \textsc{Janus}-enabled Apollo / GraphQL APIs, thus providing additional benefits in matters of reusability (\emph{NFR2}).

Second, a module is needed that specifically wraps the access control and information reduction functionality for an Apollo GraphQL server via custom directives. This is done in a separate package comprising a collection of respective directives called \textit{janus-graphql-anonym-directives} which is also available under an open source license and provided via Github and npm.\footnote{\url{https://github.com/PrivacyEngineering/janus-graphql-anonym-directives}}%

\begin{figure}[t]
    \centering
    \includegraphics[width=1.0\linewidth]{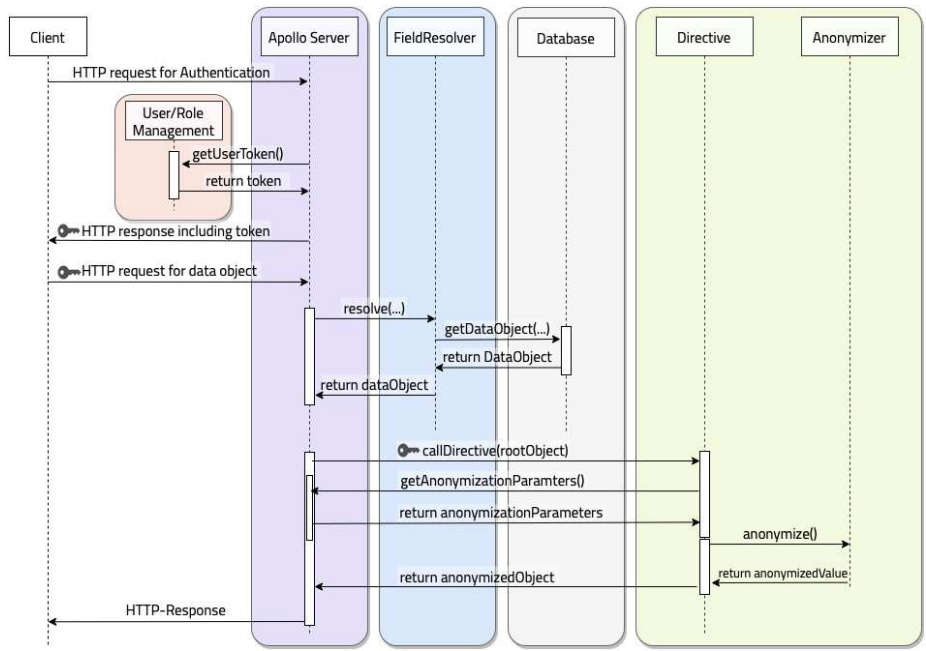}
    \caption{General sequence of operations triggered by a http request containing an exemplary data object.}
    \label{fig:architecture}
\end{figure}

Given this natural two-way split, the flow of a request for a single data object -- including the authentication and role-specific token provision exemplarily implemented in Apollo -- and the corresponding processing steps are depicted in figure \ref{fig:architecture}: %
After a http(s) request (including a previously generated role-token) is sent by the client, the respective data is fetched from the database by the resolver and returned to the server. Directives added to the corresponding data fields in the schema are called by the Apollo server directly after the resolver has fetched the data. %
The directives, in turn, use the JWT tokens to extract %
the provided role of the current requester. Depending on that role, the directive requests the role-dependent parameters %
from the developer's implementation of the directives (see section \ref{sec:usage}). %
Based on these parameters, the directive then executes the information reduction function(s)  and hands over the processed object back %
to the Apollo Server which finally returns it back to the requester. %

\subsection{Usage \& Configuration Mechanism}\label{sec:usage}

After installing \textsc{Janus} and integrating it into a given Apollo deployment, multiple custom directives for simple attribute suppression (implementing basic per-attribute access control, \emph{FR1}) and the three more advanced information reduction techniques generalization, noise, and hashing (\emph{FR2}) can be easily specified and parameterized. Noteworthily, the specification mechanism allows for role-dependent behavior as well as for the integration of components external to \textsc{Janus} itself. Besides flexible configurability (\emph{FR4}), this may also help lowering the integration overhead (cf. \emph{NFR1}), especially in complex enterprise environments.\footnote{Details on the necessary steps for installation, integration, and directive specification are provided at \url{https://github.com/PrivacyEngineering/janus-graphql-anonym-directives}.%
} A so-specified custom directive can then be added to a GraphQL schema by appending its name at the end of the definition for the data field it shall be applied to, using the common, decorator-style syntax (e.g., \texttt{@noise}) as depicted in listing \ref{lst:schema}.

\begin{lstlisting}[language=JavaScript, firstnumber=1, basicstyle=\ttfamily\footnotesize, caption={Adding the directive to the schema.},label={lst:schema}]
directive @noise on FIELD_DEFINITION
...
type Symptom {
    pain: Float @noise
    ...
}
\end{lstlisting}

\subsection{Information Reduction %
Techniques} \label{sec:reduction}
Having laid out the general architecture, implementation, and usage of \textsc{Janus}, the provided information reduction mechanisms %
shall be elaborated on in some more detail.  
The most fundamental approach for information reduction is the complete suppression of single attributes (such as names, identifiers, etc.). This functionality can be implemented through the suppression directive, which effectively implements per-attribute role based access control and can be applied to fields of any data type. In case the requester does not hold a role allowed to access a data field the directive is attached to, it simply suppresses that field's data and returns \texttt{null} instead.

Besides this basic role-based access control, \textsc{Janus} comprises three advanced information reduction methods -- generalization, noise, and hashing -- which function on different data types and can be selectively applied, independent of each other.\footnote{Details on available parameters etc. are again provided at \url{https://github.com/PrivacyEngineering/janus-value-anonymizer}.%
} %

\paragraph{Generalization:}
The generalization method supports the data types number, string, and date. Since a schema in GraphQL has explicitly typed fields, it is, however, not possible to turn a number value (integer or float) into another data type that represents a range of numbers without contradicting the schema definition. For the generalization of number values, we thus let single numbers represent a range. 
If, for example, a generalization step size of 10 is used, delimiter-based generalization results would be 0 (representing 0-9), 10 (for 10-19), 20 (for 20-29), and so forth. This ensures that exactly one number represents one range.
Because generalization is a core method for implementing data minimization, it is also implemented for dates and strings. For dates, %
natural generalization boundaries are given by the different units (second, minute, etc.). For strings, in turn, it is possible to specify how many letters should appear in plain text and at what point they should be hidden with asterisks ($*$).

\paragraph{Noise:}
The noise function %
can be used on the data types number and date. When using noise, one has to define the mathematical probability distribution to be used for sampling the noise value that is %
added to the original value. Every distribution available in the \textit{probability-distributions} package\footnote{\url{https://www.npmjs.com/package/probability-distributions}} (such as Laplacian, Normal, etc.) can be used here. Together with %
a map of configuration parameters corresponding to the distribution, developers have %
a broad flexibility in implementing the noise behavior. Since noise is mainly useful on numeric values, it is implemented for integers, floats and dates.

\paragraph{Hashing:}
Hash-based data minimization is implemented using the (so far) secure hashing function SHA3 with output lengths as available in the \textit{crypto-js}\footnote{\url{https://www.npmjs.com/package/crypto-js}} library (224, 256, 384, or 512 bits). Other hashing algorithms more resistant to brute-forcing \cite{finck-pallas-identified} as well as capabilities for including salt are not implemented at the moment but might be easily added in the future. %

\subsection{Preliminary Performance Evaluation} \label{sec:benchmark}

To experimentally determine the performance overhead caused by the integration of \textsc{Janus} and, thus, to validate the fulfillment of \emph{NFR3}, %
we prototypically implemented a GraphQL API backend for the period tracking scenario envisioned in section \ref{sec:application} and ran several performance tests against this API with and without \textsc{Janus} being active. This exemplary backend mainly consists of a GraphQL web server based on Apollo and a PostgreSQL database holding the data to be provided. The data model comprises a realistic composition of \textit{1:1}, \textit{1:n}, and \textit{n:m} relationships between entities. The data minimization directives are included as described above.\footnote{The exemplary implementation can be found here:
\url{https://github.com/PrivacyEngineering/janus-period-tracking-app}} %

For the experiments, we applied established principles of security-/privacy-related performance benchmarking \cite{pallas2020pbac}, using realistic datasets and state-of-the-art public cloud instances on Microsoft Azure in the same region. %
We measured two metrics by running respective experiments against the API instances without (\enquote{Baseline}) and with \textsc{Janus} installed: 1) the latency added by our information reduction directives and 2) the imposed reduction in matters of throughput. %
For covering aspects like general directive invocation overhead etc. separately, we also included a \enquote{no-operation} directive. In this case, requested data traversed through the directive loop before being delivered, albeit without applying any information reduction.\footnote{More details on the experiment setting are left out here due to space constraints but can be provided upon request. Employed scripts are available at \url{https://github.com/PrivacyEngineering/janus-performance-evaluation}.} %

\begin{figure}[t]
\begin{minipage}[b]{0.4\linewidth}
\centering
\includegraphics[width=0.88\textwidth]{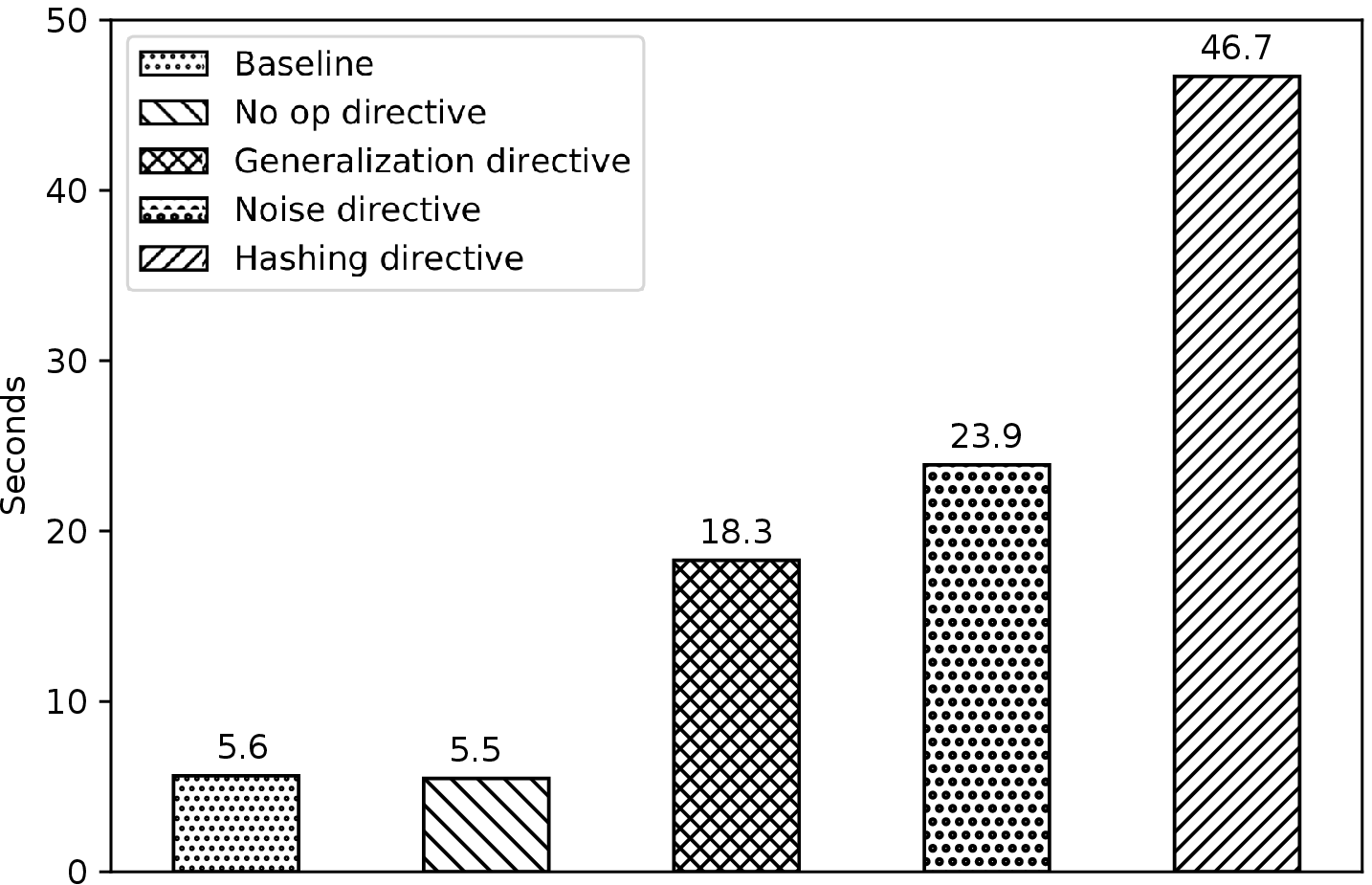}
\caption{Mean latency (seconds) with generalization, noise, and hash information reduction for 1000 data objects.}
\label{fig:barplot_directives}
\end{minipage}
\hspace{0.05\linewidth}
\begin{minipage}[b]{0.55\linewidth}
\includegraphics[width=0.86\textwidth]{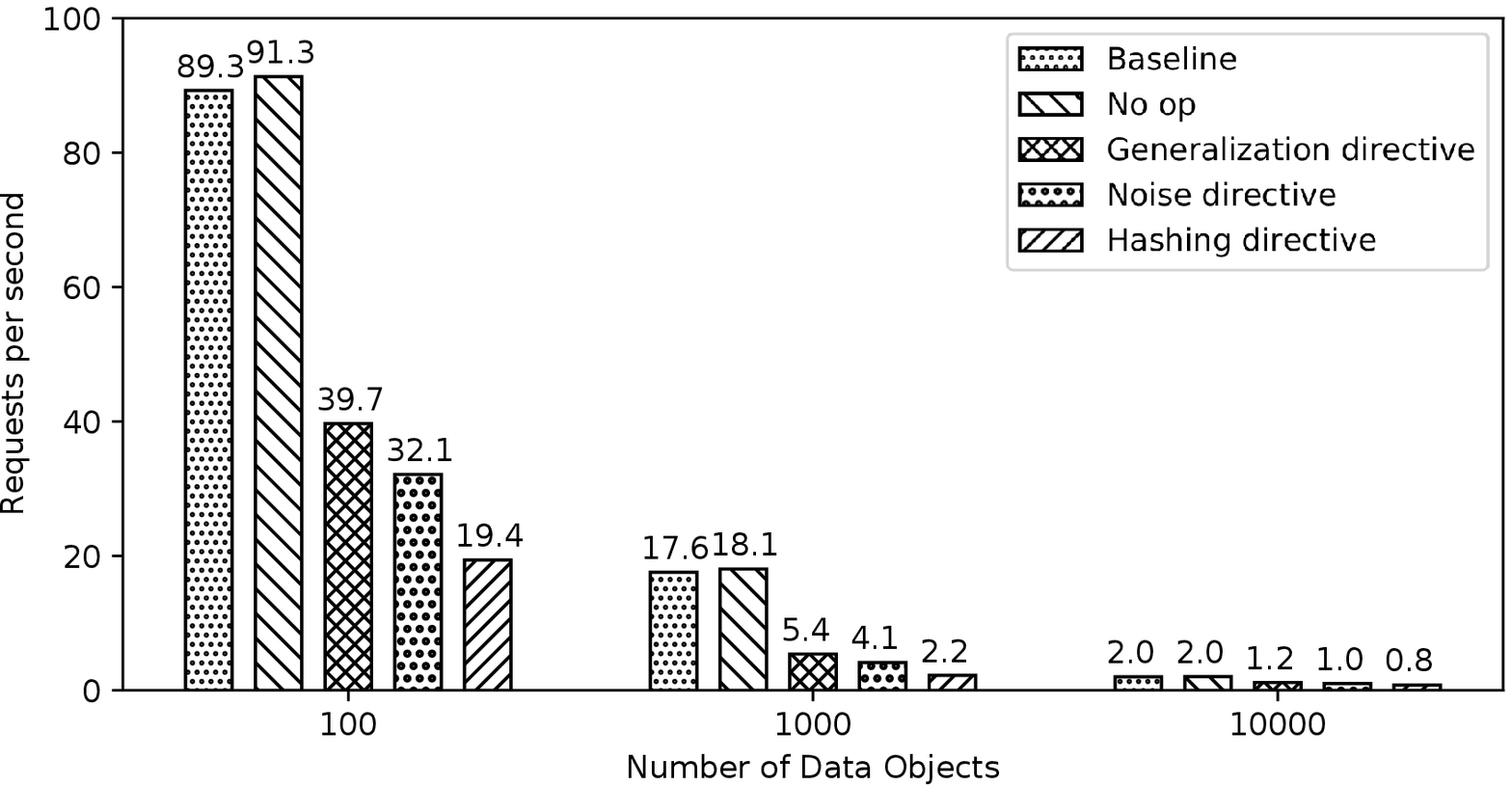}
\caption{Mean throughput (in requests per second) for different object frequencies for no operation, baseline, generalization, noise and hash information reduction methods.}
\label{fig:throughput}
\end{minipage}
\end{figure}

\paragraph{Latency:} 
Figure \ref{fig:barplot_directives} depicts the latencies observed for the different directives with 1000 data objects being requested. Noteworthily, the no-op directive does not significantly differ from the baseline, validating the efficiency of our general, directive-based  integration approach. 
When \textsc{Janus} is used with the generalization, noise and hashing information reduction methods, latencies increase at different rates: Generalization and noise lead to 3-4.5-fold latencies while hashing results in a factor of 8.5. These results are substantial, but given the comparably time-consuming operations \textsc{Janus} introduces compared to simply handing over the data from the resolver to the API endpoint, they fall within expectable and justified ranges. %
Interestingly, the relative overhead decreases significantly with more objects being requested and processed. %
For instance, moving from 1.000 to 10.000 objects results in a 8.5-fold latency for the baseline but only a 2.6-fold one with hashing being performed. We assume either some sort of static delay factor introduced by %
\textit{crypto-js} or a general, \textsc{Janus}-independent side-effect of Apollo when serializing larger responses. In any case, this aspect clearly deserves further examination in the future.

\paragraph{Throughput:} 
Figure \ref{fig:throughput} shows the measured throughput for 100, 1.000 and 10.000 data objects comparing all information reduction methods with non-operational directives and the baseline. Again, baseline and no-op do not differ significantly (with no-op in fact being slightly more performant). Of the remaining ones, generalization has the lowest impact, followed by noise and hashing. Like for the latency experiments, overheads are substantial compared to simply delivering results without any further processing, but stay within expectable and justified ranges given the computational efforts required by information reduction.
The relative loss factor again decreases with larger responses containing 10.000 data elements for all information reduction methods. This speaks in favor of above-mentioned assumption of a serialization-related side-effect.%

Given the computational overheads necessarily induced by the implemented information reduction methods compared to simply delivering data without further processing, %
these overheads are far from being unexpected. Especially for cases where data minimization %
is an indispensable precondition for implementing legally compliant API-based data provision to different parties at all, however, the observed overheads will presumably be considered reasonable (see \emph{NFR3}). %
In addition, \textsc{Janus'} concept of \emph{per-query and role-based} data minimization allows to selectively apply information reduction techniques where actually required, facilitating fine-tuned and differentiated adjustments.

Last but not least, the relative overhead of \textsc{Janus} will expectably decline in more complex application architectures involving a multitude of further factors such as data preprocessing, additional database calls, mobile data transfers, etc. (for a vivid example on the effect of such factors, see \cite{pallas2020pbac}).

\section{Limitations, Future Work \& Conclusion} \label{sec:limitations}

Introducing the concept of per-query role-based access control and information reduction to the %
domain of data-providing, query-capable Web APIs, %
identifying respective requirements, and providing a first practically usable prototype for an established GraphQL stack were the definite foci of the work presented herein. Given the rather initial state, several limitations naturally remain and various aspects had to be left open for future work.%

First of all, we refrained from adding overall complexity by complementing our approach with %
consent management mechanisms. Clearly, policy languages such as the extensive XACML \cite{anderson2003extensible} or the lightweight YaPPL \cite{ulbrichtYaPPLLightweightPrivacy2018a} would have provided more detailed and individually adjustable control over data releases. %
However, they would also require a significantly more extensive user management and custom vocabulary definitions %
and also introduce extra performance overhead. Moreover, client-side complexity would drastically increase as well, which impedes the low-effort integration of our component.

Additionally, the provision of further well-known anonymization and information reduction techniques would clearly advance \textsc{Janus'} scope of application. In particular, the integration of $\varepsilon$-differential privacy with custom additive noise mechanisms is an obvious candidate here and would allow for aggregating queries with clearly specified guarantees. Similarly, the existing information reduction methods could also be advanced with more complex functionalities like string ranges etc. %
Based on the general architecture provided herein, we expect such extensions to be %
straightforward implementation tasks based on available ready-to-use packages, which we invite the community for doing so.

Finally, more in-depth examinations of the so far only preliminarily assessed performance impacts are a clear subject for future work. For instance, the effects observed for larger responses in section \ref{sec:benchmark} deserve closer inspection, the overall impact of \textsc{Janus} in more complex end-to-end settings is so far unvalidated, the effect of combining different information reduction methods should be illuminated, etc. All this should, of course, go hand in hand with dedicated performance optimizations of the initial, unoptimized prototype implementation presented and provided herein.

These open issues notwithstanding, we herein presented the first-of-its-kind re-usable component that combines role-based access control with per-query data minimization for modern Web and in particular GraphQL APIs. The application scenarios for \textsc{Janus} in the course of practical privacy engineering are manifold and were illustrated with a period-tracking application example without limiting its generality for other real-world scenarios. The performance impacts were shown to be non-negligible but still appear to be acceptable, especially for usecases that would not be implementable in a legally compliant way without solid information reduction being applied before the release of initially personal data. %

We emphasize the particular importance of both legal and technical requirements that guided the design of \textsc{Janus'} general architecture and its implementation. Under these practice-oriented guidelines, we are convinced of the possible low-effort integratability of our open source component into many real-world systems for effectively heightening the level of privacy. Noteworthily, \textsc{Janus} does not, per se, provide any guarantees in matters of regulatory compliance, given the multitude of factors influencing what needs to be implemented in a particular case. It does, however, equip developers of data-providing GraphQL APIs with the technical capabilities for fulfilling case-dependent legal data minimization obligations in a handy, highly configurable, and easy to use manner. %